# THE DYNAMIC SPACE OF GENERAL RELATIVITY IN SECOND ATOMIZATION

*Lukas A. Saul*
University of New Hampshire, Dept. of Physics & Institute
of Earth, Oceans and Space

## Abstract

The notion that the geometry of our space-time is not only a static background but can be physically dynamic is well established in general relativity. Geometry can be described as shaped by the presence of matter, where such shaping manifests itself as gravitational force. We consider here probabilistic or atomistic models of such space-time, in which the active geometry emerges from a statistical distribution of 'atoms'. Such atoms are not to be confused with their chemical counterparts, however the shift of perspective obtained in analyzing a gas via its molecules rather than its bulk properties is analogous to this "second atomization". In this atomization, space-time itself (i.e. the meter and the second) is effectively atomized, so the atoms themselves must exist in a 'subspace'.

Here we build a simple model of such a space-time from the ground up, establishing a route for more complete theories, and enabling a review of recent work. We first introduce the motivation behind statistical interpretations and atomism, and look at applications to the realm of dynamic space-time theories. We then consider models of kinetic media in subspace compatible with our understanding of light. From the equations governing the propagation of light in subspace we can build a metric geometry, describing the dynamic and physical space-time of general relativity. Finally, implications of the theory on current frontiers of general relativity including cosmology, black holes, and quantum gravity are discussed.

## I. Introduction – the Atom

*In which the ideas of atomism are reviewed and revived for a modern approach.*

The ancient ideas of atomism, as attributed to Democritus and Leucippus, are in some ways a fundamental analysis of human perception of reality. Objects are made up of the sum of their parts, and the parts are in turn made of the sum of their parts. The original idea carried this recursion to a possible end: the *Atom*, which has no parts. Although the word itself comes from the Greek "uncuttable", we will rename this 'indestructible' portion of the idea to be compatible with modern terminology:

> **Strong Principle of Atomism:** *There exist fundamental entities which cannot be subdivided (which have no parts), and which cannot be destroyed.*

It is tempting take on the philosophical argument of the possibility of an object with no parts, or to discuss the implications including an infinite past and future. However, some utility of atomism comes from a corollary to this principle, which is that the atoms are extremely small and numerous. To be more specific, the macroscopic behavior of a system does not depend on the characteristics of an individual atom. This important concept does not require perfect 'uncuttable' behavior to still be of some use. We can define a weak principle as well, compatible with the atoms as championed by Boltzmann and Dalton:

> **Weak Principle of Atomism :** *In a given system of a certain size and time scale of interest, there exist constituent entities of sufficient number for which specific knowledge of an individual's properties is not necessary for predicting the behavior of the original system.*

In this probabilistic interpretation, the atoms have become 'molecules', for which we need not know the properties of any single one, rather averages of many. Molecules do not of course adhere to the strong principle of atomism, for they are made of elements and more fundamental entities (parts), and can be broken apart with electrical or chemical reactions. However, many of the advantages of the atomic approach still hold. We can describe for example temperature and heat of a system in terms of the average kinetic energy its molecules, and quantitatively model thermodynamics, sound transmission, and other phenomena. If we describe a system in such a manner we must be careful to include in our equations any processes which might be sources or sinks of our 'weak' atoms,to make up for the fact that they are not truly 'uncuttable'.

This weak principle allows for all kinds of new 'atoms', e.g. the individual as an atom of society, the cell as an atom of an organism, or a star as an atom of a globular cluster. However, it still might offend the staunchest empiricists in some cases, when a set of experimentally determined physical laws seems easier to use than an axiomatic theory. In efforts to keep such naysayers with us a bit further through this discussion we offer a final and somewhat gutted version of atomism:

> **Weak Principle of Atomism (2):** *For a given system of a certain size and time scale of interest, the evolution can be described as though there exist constituent entities of sufficient number for which specific knowledge of their respective constituents is not necessary for predicting the behavior of the system.*

In this scenario, an observer need not accept the physical existence of an atom at all. Rather, one must only accept that the observables of a system can be described with the statistical mechanics.

For the purposes of this report, we will not argue the merits of one of these principles over the other, though such discussion is extremely interesting from a philosophic or metaphysical viewpoint. We will only discuss one application of the principles of atomism, to describing the general-relativistic quantum vacuum, or physical space-time. The reader is free to choose which principle of atomism is most applicable to physical space-time, the methods and equations described herein being applicable under any of them.

## II. Kinetic Theory of Space-Time

*In which the dynamical equations of space-time atoms are introduced, with a simple trial model of a phase space, motivated by physical symmetries of quantum mechanics and electromagnetic theory.*

A goal here is to take as our system an "empty" space, i.e. a quantum vacuum in a general relativistic space-time, and describe its properties in a statistical atomic framework. Such a procedure is similar to that laid out by early atomists, or to the mechanical models of Maxwell or the Monadology of Leibniz. It has been discussed recently by [Meno, 1991], who uses the term "gyrons" to describe the atoms of space-time, and suggests that these entities exist in numbers as great as $10^{90}$ per cubic centimeter. Quantum mechanics has also been described using such statistical-kinetic (atomistic) techniques to describe the vacuum [e.g. Winterberg, 1995; Kaniadakis, 2002]. The technique has also been applied to general relativity, for example in the approach of "loop quantum gravity" [e.g. Smolin, 2004], who suggested a similar number density to that of Meno. The kinetic theory has also been proposed as a road to quantum gravity [Hu, 2002]. The word "kinetic" is used because we borrow formalism from the kinetic theory of gases to model our new system, and because we allow in general non-equilibrium distributions to exist.

### A. "Subspace"

The basic tenet of describing the constituents of space-time as atoms runs into one major question which is unique to this application of atomism. What space do the atoms exist in themselves, if the space we know is made of their collective properties? The answer is that we must posit a *subspace*, that space in which the atoms of our space-time reside. It is important to realize that the term as used here is removed from its set theory definition, in which it would designate a space contained by another. In fact the term 'superspace' may be more appropriate in that context, as the subspace dynamics contains the dynamics of physical space. However, the analogies with the 'subspace' of science fiction as well as the much smaller size scale of the constituent particles elicit use of the term, at least for this report.

There are many philosophical issues which arise from such a theory that are not addressed here. Some might suggest that the assumption of a subspace is an artificial construct, which negates the advantage of simplicity that the kinetic theory introduces. In the theory however, this subspace is more fundamental than the usual space-time we are familiar with. One could argue that the physical space we are familiar with is actually the more artificial construct. The question of what makes up the subspace will not be dealt with here, other than to say that its answer lies partly in which principle of atomism one chooses to apply, and at what scale.

In some ways this is a theory of "pre-geometry", in that it attempts to derive a geometry (see e.g. Wheeler in [Marlow, 1980]). However, rather than deriving all of geometry, what we are really doing is "passing the buck". Although we can derive our physical space-time geometry from the subspace dynamics, we have pulled the subspace geometry from thin air (by hypothesizing its existence), and so the same questions as to its underpinnings exist as did originally.

The assumption of the existence of an abstract mathematical space may seem at first extravagant, unnecessary, or even unverifiable. However, physicists for the past century have been pursuing similar approaches, using for example infinite dimensional Hilbert spaces, cosmological five dimensional manifolds, or even more abstract mathematical spaces such as p-branes and orbifolds, to explain physics. Modern string theories study conformal maps from higher dimensional spaces to physical space; we pursue a similar route here demonstrating how to map from subspace to physical space using a kinetic approach. Strings also appear in vortex dynamics, which could arise from a kinetic space-time model. In these ways the atomic theory can act as a unifying tool, allowing diverse other theories to be classified in terms of their interpretation as subspace kinetics.

## 1. Distribution Functions in Subspace

Formalism can be developed for the general case of atoms in a subspace manifold. Each atom is assumed to have some characteristics, i.e. a 'location' in a phase space. The dimensions of such a phase space are not limited to spatial dimensions, but include any other degree of freedom that can be envisioned. This could include anything from hypercomplex spaces to Grassman algebras of supersymmetry. In this scenario a group of atoms are completely macroscopically described by the distribution functions $f_i(\vec{\alpha}_j)$, which represent the density of atoms of type $i$ to have properties (locations) $\vec{\alpha}_j$. More specifically, the integral of the distribution function over some phase space volume $\prod_j \Delta\vec{\alpha}_j$ gives the number of atoms in that volume. Macroscopic physical properties must then be computed as statistical averages of this distribution, and their evolution computed with a Hamiltonian or other dynamics. For the case of a phase space $\vec{\alpha} = \vec{p}, \vec{q}, t$ with generalized position and momentum vectors $p$ and $q$ of arbitrary dimension (and holonomic constraints), a Hamiltonian $H_i(\vec{p}, \vec{q}, t)$ describes the motion of constituents of type $i$ in the distribution. The distribution function itself evolves as:

$$\frac{\partial f_i}{\partial t} = [H_i, f_i] \tag{1}$$

where we have used the Poisson bracket notation,

$$[a,b] = \frac{\partial a}{\partial q_i}\frac{\partial b}{\partial p_i} - \frac{\partial a}{\partial p_i}\frac{\partial b}{\partial q_i} \tag{2}$$

These equations, derived from the Liouville theorem of conservation in phase space, fully describe the evolution of any system with Hamiltonian dynamics. They are rigorously derived for statistical distributions, which themselves can be made rigorous using the concept of generalized functions. The interested reader should consult one of numerous reviews of kinetic theory [e.g. Liboff, 1969; Wu, 1966]. They are also applicable to numerous other systems that can be treated kinetically, e.g. [Marmanis, 1998a].

With an infinite number of possible subspaces, atomic characteristics (phase spaces), and dynamics to choose from, it is difficult to know where to begin. To guide this search we will consider a simple class of atomistic space-times, based on a single type of effectively indestructible atom in a force free Euclidean three dimensional space plus time. Somewhat surprisingly, this choice yields diverse kinetic phenomena capable of describing many of our physical observables. Later we can return and discuss what other choices for a subspace yield useful physics. We note in passing that it is likely that many different choices of subspace dynamics could yield physical reality as we observe it. In this case the physicist is free to choose his favorite, using Occam's razor or another principle to guide the decision.

## 2. Higher Dimensional Phase Spaces in Euclidean Subspace

One simple model of the atom is as a point which moves through a three dimensional subspace. The only properties such an atom can have are that of position and velocity. The distribution function describing a large group of these point atoms has the form $f = f(\vec{r}, \vec{v}; t)$, which represents a seven dimensional phase space. This choice works well when describing for example some monatomic gases or plasmas atomically.

By giving the atoms some finite volume or shape we also can include orientational and rotational freedom in the phase space. In such a manner we can visualize higher dimensional phase spaces, still using only a three dimensional Euclidean configuration space. Examples of resultant distributions functions are shown in Table 1:

| Distribution Function | Dimensions of Phase Space | Atomic Properties |
|---|---|---|
| $f = f(\vec{r}, \vec{v}; t)$ | 7 | Position, velocity. |
| $f = f(\vec{r}, \vec{v}, \vec{\alpha}, \vec{\omega}; t)$ | 12 | Position, velocity, orientation, spin. |
| $f = f(\vec{r}, \vec{v}, \vec{\omega}; t)$ | 10 | Position, velolcity, spin. |

For subspaces with three configuration dimensions, each vector is expressed as three scalar components, except for the orientation $\vec{\alpha}$ which is described by two scalars.

Higher dimensional manifolds have long been proposed as necessary for unification efforts, starting with pioneering efforts of Weyl (1918), Kaluza (1921), and Klein (1926), discussed and reprinted in [O'Raifeartaigh, 1997]. In the kinetic approach these "extra dimensions" can appear in a more easily understandable way, i.e. in kinetic phase spaces of three dimensional configuration spaces.

## B. Basic Equations of Kinetic Theory

We consider here the simple kinetic model in subspace with a ten dimensional phase space. The goal is to see how macroscopic laws of physics emerge and adhere to our observations and experiments, so that others can return with more sophisticated models as necessary. We posit the existence of a single type of atom, existing in Euclidean $\mathbb{R}^3$ plus time. Motivated by requirements of the theory to produce quantum and electromagnetic interactions [see e.g. Meno, 1991; Recami, 1998; Saul, 2003] we include the additional characteristic of a spin degree of freedom for our atoms. We follow here in part the notation of [Saul, 2003]. First we will develop the kinetic equations, then return and identify how they apply to physical observables. The approach will serve as a review for those familiar with the kinetic derivation of hydrodynamics, fluid mechanics, or magneto-hydrodynamics, while serving as a sample algorithm for the enterprising reader wishing to test his own model of subspace.

### 1. Boltzmann Equation and Conservation Equations

In a detailed series of papers [Curtiss, 1957], a modified Boltzmann equation and transport equations for a distribution like our simple model were derived, using the distribution $\tilde{f} = \tilde{f}(\vec{x}, \vec{v}, \vec{\alpha}, \vec{\omega}; t)$, in which particles have positional, translational, orientational, and rotational freedom in time. In our case the coordinates in these equations are in arbitrary subspace units, as we have not yet introduced a physical space-time metric. Here we use a slightly simplified distribution function, assuming uniformity in the orientational distribution of the particles, i.e. assuming for now that a distribution function in a ten dimensional phase space $f = f(\vec{x}, \vec{v}, \vec{\omega}; t)$ fully characterizes the system. Here $\vec{x}$ is the position, $\vec{v}$ is the velocity, and $\vec{\omega}$ is the angular velocity of a constituent particle, at a time $t$. We can also view this simplification as the more general distribution of [Curtiss, 1957], integrated over all orientations $\vec{\alpha}$.

The distribution function gives the positive ten dimensional phase space density of particles with these properties. The Boltzmann equation describing its evolution is (no external forces):

$$\boxed{\frac{\partial f}{\partial t} + \frac{\partial f}{\partial x_i} v_i = \left\{\frac{\partial f}{\partial t}\right\}_{col.}} \quad (3)$$

This equation comes from the general form in (1), with a judicious separation of the Hamiltonian. The Hamiltonian is treated as a free streaming (no external force) piece, along with a piece that acts only over very short times and is associated with particle collisions. The collisional piece of the Hamiltonian gives the right hand side of equation (3). Much work has gone into making this equation rigorous; proving its validity, and quantifying the collision term, often including external forces as well. Unfortunately the details cannot be covered here, but again interested readers are directed to good reviews by e.g. [Liboff, 1969; Wu, 1966]. For our purposes now equation (3) will serve as a master equation, allowing us to derive dynamical equations of statistical moments. For example, we can integrate (1) over all phase space (in our case $d^6\Gamma$ or $d^3\vec{v}d^3\vec{\omega}$), to obtain the usual continuity equation:

$$\frac{\partial n}{\partial t} + \frac{\partial}{\partial x_i}(n\bar{v}_i) = 0 \tag{4}$$

where we have assumed that the particles are neither created nor destroyed in collisions, and we have defined the number density $n(\vec{x},t) = \int f d^6\Gamma$. We can also multiply (3) by $\vec{v}$ or by $\vec{\omega}$ before integrating, giving the transport equations:

$$\frac{\partial}{\partial t}(n\bar{v}_i) + \frac{\partial}{\partial x_j}(n\langle v_i v_j \rangle) = -E_i \tag{5}$$

$$\frac{\partial}{\partial t}(n\bar{\omega}_i) + \frac{\partial}{\partial x_j}(n\langle \omega_i v_j \rangle) = F_i \tag{6}$$

The notation is that a moment: $\langle Q \rangle = \frac{1}{n}\int Q f d^6\Gamma$ is a function of space and time only, for any function $Q$ of the properties of an individual particle. In the atomic approximation (weak or strong) it is the moment $\langle Q \rangle$ rather than an individual atom's $Q$ which is the macroscopic observable of interest. For ease of notation we write averages $\bar{v}_i = \langle v_i \rangle$ and $\bar{\omega}_i = \langle \omega_i \rangle$. We have also defined two vectors:

$$-E_i = \left\{\frac{\partial n\bar{v}_i}{\partial t}\right\}_{col} = \int \left\{\frac{\partial f}{\partial t}\right\}_{col} v_i d^6\Gamma \tag{7}$$

$$F_i = \left\{\frac{\partial n\bar{\omega}_i}{\partial t}\right\}_{col} = \int \left\{\frac{\partial f}{\partial t}\right\}_{col} \omega_i d^6\Gamma \tag{8}$$

representing the collision terms in the above conservation equations. In non rotational media with fully elastic collisions, $E$ will vanish, as linear momentum is neither created nor destroyed in the collisions. Equations (5) and (6) can also each be combined with the continuity equation to change the form to the well known Navier-Stokes equations (for bulk velocity) and a Navier-Stokes analog for the bulk spin (not shown here).

It is useful to represent the velocity and rotation of an individual particle as a sum of the bulk or average speed at that location plus any 'peculiar' motion, i.e.: $v_i \equiv \bar{v}_i + V_i$ and $\omega_i \equiv \bar{\omega}_i + \Omega_i$ (capital letters represent the peculiar motion). The average of these peculiar motions is zero by definition, i.e. $\langle \Omega_i \rangle = \langle V_i \rangle = 0$. Note that we are not assuming anything about the size of these quantities here (the peculiar velocity or spin can be larger than the bulk), only defining new variables. The convection terms from (5) and (6) can now be rewritten, and pressure tensors defined:

$$n\langle v_i v_j \rangle = n\bar{v}_i \bar{v}_j + n\langle V_i V_j \rangle \equiv n\bar{v}_i \bar{v}_j + p_{ij} \qquad (9)$$

$$n\langle \omega_i v_j \rangle = n\bar{\omega}_i \bar{v}_j + n\langle \Omega_i V_j \rangle \equiv n\bar{\omega}_i \bar{v}_j + s_{ij} \qquad (10)$$

This spin pressure tensor $s_{ij} = n\langle \Omega_i V_j \rangle$ will vanish if the distribution function is isotropic in either $\vec{V}$ or $\vec{\Omega}$ as the integrand will be odd in these variables. The off-diagonal components of the pressure tensor $p_{ij} = n\langle V_i V_j \rangle$ will similarly vanish for f isotropic in $\vec{V}$ space. Also, the collision terms for the velocity equation and the angular velocity equation $\vec{E}$ and $\vec{F}$ will vanish if $\left\{ \dfrac{\partial f}{\partial t} \right\}_{col}$ is isotropic in $\vec{V}$ or $\vec{\Omega}$ respectively. In this case by "isotropy in $\vec{V}$ space" we mean that the distribution function depends only on the magnitude of $\vec{V}$ and not on its direction.

Many important analyses of fluid mechanics rest on evaluation of the pressure tensor. For example, the pressure tensor in an incompressible Newtonian fluid is [Landau & Lifshitz, 1959]:

$$p_{ij}(\vec{x},t) = \delta_{ij} p(\vec{x},t) - \eta \left( \frac{\partial \bar{v}_i}{\partial x_j} + \frac{\partial \bar{v}_j}{\partial x_i} \right) \qquad (11)$$

where $\eta$ is the viscosity, and $p(\vec{x},t)$ is the scalar pressure. We say the pressure tensor is diagonal if it takes the form (11) with zero viscosity; we say the pressure is isotropic if it is independent of the spatial coordinates.

## 2. Wave Equations

Rather than having to solve for the moments such as density, bulk velocity, bulk spin, or pressure tensor explicitly as functions of space and time, we can learn much about the behavior of our medium by examining the propagation of disturbances or oscillations. The first oscillations usually discussed in the kinetic theory of gases are compressive or 'sound' waves, so we describe them here as well by way of an introduction.

Equations (4) and (5) can be combined into one, by taking the time derivative of the first and the spatial gradient of the other, yielding:

The Dynamic Space of General Relativity in Second Atomization                 161

$$\frac{\partial^2 n}{\partial t^2} = \frac{\partial}{\partial x_i}\frac{\partial}{\partial x_j}(n\langle v_i v_j\rangle) + \frac{\partial E_i}{\partial x_i} \qquad (12)$$

This equation governs compressive or sound-like waves. It reduces to the ordinary form of the wave equation:

$$\frac{\partial^2 n}{\partial t^2} = c_s^{\,2}\nabla^2 n \qquad (13)$$

for $E_i$ vanishing or constant, the pressure tensor diagonal and isotropic, and the bulk speed vanishing or constant. If the number density is the same everywhere, there is no power in these wave modes.

Another wave equation emerges from transport of the spin pressure. Multiplying the master equation (3) by the tensor $v_i\omega_j$ and integrating over phase space yields a higher order transport equation:

$$\frac{\partial}{\partial t}(n\langle \omega_i v_j\rangle) + \frac{\partial}{\partial x_k}(n\langle \omega_i v_j v_k\rangle) = G_{ij} \qquad (14)$$

where the collision term $G_{ij}$ is a tensor defined analogously to the vectors $\vec{E}$ and $\vec{F}$. The rank three tensor moment in the second term is the *spin pressure transport tensor*.

We can now take the derivative of (14) with respect to $x_j$, and the derivative of (6) with respect to time, and combine these two equations, giving:

$$\boxed{\frac{\partial^2(n\bar{\omega}_i)}{\partial t^2} = \frac{\partial}{\partial x_j}\frac{\partial}{\partial x_k}(n\langle \omega_i v_j v_k\rangle) + \frac{\partial F_i}{\partial t} - \frac{\partial G_{ij}}{\partial x_j}} \qquad (15)$$

At this point we have made no assumptions of the ten dimensional distribution function as defined, other than its statistical applicability. In other words, equation (15) is as general as possible for the simple "orientropic" subspace model we have chosen to consider. We will see that this equation is extremely important in space-time dynamics; indeed it will allow us to define physical space-time by modeling electromagnetic radiation.

To see why equation (15) is of any immediate use requires us to consider a specific distribution. Assume appropriate isotropy of the distribution function collision term so that the source terms, the vector F and tensor G, vanish. Take a frame of reference moving with the bulk flow so that $\bar{v}_i = 0$ for i=0,1,2. Equation (13) can now be written:

$$\frac{\partial^2 (n\bar{\omega}_i)}{\partial t^2} = \frac{\partial}{\partial x_j}\frac{\partial}{\partial x_k}\left[ n\bar{\omega}_i \langle V_j V_k \rangle + n\langle \Omega_i V_j V_k \rangle \right] \tag{16}$$

If we further assume the distribution function to be isotropic in peculiar angular velocity $\bar{\Omega}$, the final moment will vanish. We can also assume constant density in space and time and cancel $n$ from this equation, giving the form:

$$\frac{\partial^2 \bar{\omega}_i}{\partial t^2} = \frac{\partial}{\partial x_j}\frac{\partial}{\partial x_k}\left[ \bar{\omega}_i \langle V_j V_k \rangle \right] \tag{17}$$

At this point we have a familiar wave equation. A diagonal pressure tensor $\langle V_j V_k \rangle = c^2 \delta_{jk}$ gives the form

$$\frac{\partial^2 \bar{\omega}_i}{\partial t^2} = c^2 \nabla^2 \bar{\omega}_i \tag{18}$$

with d'Alembert solutions propagating at the speed $c$. Such a wave equation governs propagation of bulk spin perturbations. Presumably wave power would be added to these modes by disturbances even in incompressible fluids. The difference between equations (13) and (18) is that in the latter the fluctuating quantity is a vector $\bar{\omega}_i$, rather than the scalar density $n$. This enables a mechanical model of polarized transverse wave propagation.

## C. Electromagnetism

Thus far kinetic equations have been examined governing the evolution of atoms in a Euclidean phase space including spin (subspace). Nothing has been said about identifying the bulk properties of such distributions with physical observables, or how such a subspace will reproduce physical space. However, the motivation in choosing the phase space as we have done is largely from the symmetry with the known properties of electric and magnetic fields. In fact many of the diverse phenomena of electromagnetism can be explained from such an atomistic viewpoint, as pointed out by Maxwell in his treatise "On physical lines of force". Maxwell's equations, and even the Lorentz force, can emerge naturally from the first transport moment equations even without including the spin phase space freedoms of our simple model. We will not review here the details or the questions that emerge in this interpretation here, as our main goal is explaining the curvature or gravitational properties of space-time.

We will try to model the dynamics so that under the proper conditions, the coordinates of subspace map directly to physical space-time coordinates. Only in the presence of strong fields or relativistic motion will curvatures in the mapping manifest themselves, and indeed these manifestations are force fields. For the purposes of this paper we will not consider detailed implications of atomistic theories on electromagnetism. This topic is extremely interesting and well introduced in the literature, especially in the two volume text [Whittaker, 1960]. For a good modern review see the gyron interpretation of [Meno, 2000] or the metafluid approach in [Marmanis, 1998b]. We will need to be specific here about only one electromagnetic phenomenon which is a cornerstone of special and general relativity: light.

In view of our goal of simplicity we use now only the single simple model for light as described in equations (16-18), the spin wave equation. While other atomistic models are certainly possible, and may be necessary to model the diverse phenomena of quantum physics and general relativity, it will be enlightening to complete the procedure with one model, to outline the application. We take as a starting or grounding point conditions in which the *x* and *t* subspace coordinates in equation (16) match exactly real spatial and temporal variables.

## III. Physical Space and Metric

*In which physical space is defined and the quasi-metric introduced via the optical or analogue interpretation of general relativity.*

We will consider real or physical space-time to be a local special relativistic set of coordinates. In this physical space-time, light travels 299,792,458 meters in a second in vacuum (any vacuum!) and 9,192,631,770 oscillations of a Cesium-133 atom take 1 second. The truly objective reader may not be happy with this definition. Why are these arbitrary units, defined from the phenomenon of electromagnetic waves and atomic electrons, given the godly status of definers of physical or real space? The answer is in our faculty. Being built of electromagnetic forces and atomic chemistry, we are biased toward such a space. The official meter stick that once defined our meter and therefore our measure of space was held together by electromagnetic forces, and so was a similar convention. Although there is no "ultimate preferred physical coordinate" or absolute space, the electromagnetic choice of units is a natural one for the human condition, enjoys international consensus for use, and so will be referred to here as real or physical space-time.

It is not immediately clear that this "locally special relativistic" coordinate set will be uniquely defined from the subspace dynamics. It is likely in the very least that certain diabolical cases of subspace dynamics will render the above definition of physical space time via the meter and light untenable. However, as long as we can recreate observable space-time from some subspace dynamics, our theory is a success, and the diabolical cases can be considered later on their own merit. Again we forge ahead, ignoring for now the many forks in the road before us.

### A. Metric and Quasi-Metric

We are concerned now with the details of how distances and times are measured. Mathematicians have considered the finer details of such schemes, in what is known as measure theory. The precise definition of a distance measuring scheme is known as a 'metric', which allows the assignment of a scalar quantity to any two points in space, i.e. the distance between them. We have loosely defined such a metric above (in three dimensional space), by using the transmission of light in vacuum between two such points as the metric. We have also assumed the existence another subspace metric (Euclidean, for the case of our simple model) for description of subspace dynamics. However, to map these manifolds effectively to physical space-time we will require more mathematical formalism.

### 1. Special Relativity

Although the choice of metric using light may seem natural to those of us who grew up with the modern dictionary definition of 'meter', there are major complications to this choice. This choice requires that the speed of light be a constant (in vacuum). If we require certain equivalence of inertial frames this leads to the modified dynamics of special relativity, including the Lorentz transformation of coordinates, and such phenomenon as mass, length, and time dilation, as pointed out by Einstein a century ago. To make a long story short, the special relativistic geometry or Minkowski space can be described with an 'invariant space-time interval':

$$ds^2 = dx^2 + dy^2 + dz^2 - c^2 dt^2 \tag{19}$$

The statement that this interval $ds^2$ is invariant under coordinate velocity transformations uniquely determines the Lorentz transformations and its associated group. In fact, the mathematical entity $ds^2$ of (19) is extremely useful, describing the geometry of our coordinate space, and the coordinates in that equation can be called "physical", i.e. the $x_i$ can be measured in meters and $t$ can be measured in seconds. The form is similar to that for a general quadratic metric on a four dimensional manifold of points $x^\mu$, which gives as the infinitesimal distance $ds$ from a point with coordinates $x^\mu$ to a point with coordinates $x^\mu + dx^\mu$:

$$ds^2 = g_{\mu\nu} dx^\mu dx^\nu \tag{20}$$

However, the invariant space-time interval of (19) does not satisfy two of the most fundamental properties of real metrics: that they be positive definite, and that if the distance between two points is zero they are the same point. None the less, the 3-space metric is contained in this expression, for example by choosing a "spacelike slice", or setting $dt = 0$. The 3-space metric in special relativity is not frame independent, and so the physics is best encapsulated in the form of (19). For these reasons it is referred to as a *quasi-metric*. It is worth pointing out that the quasi-metric (19) is not a unique invariant. Of course any function of this $ds^2$ is also invariant under changes of inertial reference frame and can equally represent physical geometry.

We will not review the extensive theoretical and experimental work on special relativity here, other than to note its reliance on the constancy of light speed as an axiom with which to define metric and establish coordinate bases. This choice requires the wave equation governing light to be invariant in inertial frames. However, the flat space wave equation (e.g. equation (18)) is not invariant under Galilean transformations. The coordinate transformations that leave the wave equation invariant turn out to be what are now called the Lorentz transformations, derived in this way as early as 1890 by Voigt. We will use this approach to turn more general forms of a wave equation into more general quasi-metrics.

## B. The Optical Analogy and General Relativity

Because the theory of special relativity (SR) is built on light, it should come as no surprise that much work has gone into interpretations of general relativity (GR) also based solely on the propagation of wave disturbances. A number of optical or other analogue approaches to GR have been suggested [e.g. Nandi & Islam, 1995 and references therin] and an international conference on the topic was held in Brazil in 2000. Recent interest has been sparked by the analogy of light propagation near black holes to properties of other wave propagation in other singular media [e.g. Visser, 1998]. The approach is not that the curved space-time field of GR is created or shaped by light, as in quantum field theories, but that light is an effective tracer or probe of this curvature, as well as a potential modeling tool [e.g. Visser, 2002; Unruh, 2003]. We pursue here the approach of deriving the quasi-metric invariant from the wave equation, without calculating the associated index of refraction that is often the cornerstone of the optical approach.

### 1. Equations of State

We will begin with equation (15) in the absence of bulk flow, and neglecting the source or collision terms for now (or assuming their isotropy):

$$\frac{\partial^2 (n\bar{\omega}_i)}{\partial t^2} = \frac{\partial}{\partial x_j} \frac{\partial}{\partial x_k} (n\langle \omega_i V_j V_k \rangle) \tag{21}$$

This is not technically a wave equation until an equation of state has been adapted to associate the argument of the spatial derivatives with the argument of the time derivatives, i.e.

$$n\langle \omega_i V_j V_k \rangle = f_{ijk}(n\bar{\omega}_i) \tag{22}$$

(no implied summation). We are assuming that the fluctuations of the tensor $n\langle \omega_i V_j V_k \rangle$ are solely functions of the perturbation $(n\bar{\omega}_i)$. It is possible to assume constant number density of our atoms and divide through by $n$. However, we will leave this density in the equations for now, as the same character emerges when examining fluctuations of the bulk spin alone.

The obvious undisturbed or vacuum form of this equation of state (22), that gives us the quasi-metric (19) and flat space SR is:

$$n\langle \omega_i V_j V_k \rangle = c^2 n\bar{\omega}_i \delta_{jk} \tag{23}$$

This gives us equation (18), the standard wave equation, but leaves no room for further curvatures or freedoms in our quasi-metric.

To introduce such freedom we consider small deviations $\eta_{jk}$ from the flat space equation of state:

$$n\langle \omega_i V_j V_k \rangle = n\bar{\omega}_i(c^2\delta_{jk} + \eta_{jk}) \tag{24}$$

Plugging this into equation (15) gives us:

$$\frac{\partial^2(n\bar{\omega}_i)}{\partial t^2} = c^2\nabla^2(n\bar{\omega}_i) + \eta_{jk}\frac{\partial^2(n\bar{\omega}_i)}{\partial x_j \partial x_k} \tag{25}$$

where $\eta_{jk}$ have been assumed to have very small spatial derivatives. This form must be invariant in all inertial frames if we are using physical space-time coordinates, and so the associated invariant quasi-metric looks to be of the form:

$$ds^2 = d\bar{x}^2 + \frac{\eta_{jk}d\bar{x}^2 dt^2}{dx_j dx_k} - c^2 dt^2 \tag{26}$$

This form reduces to the usual line element (19) when the perturbation is zero, and for a diagonal perturbation $\eta_{jk} = \eta\delta_{ij}$ becomes

$$ds^2 = d\bar{x}^2 - c^2(1-\eta)dt^2 \tag{27}$$

which yields Newtonian gravity for a perturbation $\eta = \frac{2MG_N}{c^2 r}$ [see e.g. Misner, Thorne and Wheeler, 1970; Dirac, 1975]. The optical approach can also duplicate the geometry of the Schwarzschild solution, which is the spherically symmetric solution of Einstein's equations, as showed by [Nandi & Islam, 1995]. In that work the Schwarzschild quasi-metric was derived from an ad-hoc definition of the index of refraction, whereas in our current approach the Schwarzschild geometry emerges from an ad-hoc assignment of a two-component non-diagonal pressure tensor (equation of state) [Saul, 2003]:

$$\langle V_i V_j \rangle = \begin{matrix} c^2 & 0 & 0 \\ 0 & c^2(1-\eta) & 0 \\ 0 & 0 & c^2(1-\eta) \end{matrix} \tag{27}$$

here the first column and row of the matrix represent the radial coordinate (we have moved to spherical coordinates), and the resultant Schwarzschild quasi-metric is:

$$ds^2 = dr^2(1-\eta)^{-1} + r^2 d\phi^2 + r^2\sin^2\theta d\theta^2 - c^2(1-\eta)dt^2 \tag{28}.$$

## 2. Eikonal Approximation

We can be more rigorous in our derivation of the quasi-metric tensor. The association of this geometry with light rays as defined by a wave equation is known as geometrical optics. There exist a few routes to describing light rays or geodesics from the fundamental wave equation, which is not a trivial enterprise in inhomogeneous or anisotropic media. Some such methods include treating the rays as always perpendicular to the Poynting vector, defining an associated minimized action or Hamiltonian (Hamilton-Jacobi approach), and the so called eikonal approximation. These methods are reviewed in e.g. [Born and Wolf, 1969; Stavroudis, 1972]. We consider here the eikonal approximation as it applies to our modified wave equation for our simple model.

In the eikonal approximation the fluctuating field (here $n\bar{\omega}_i$) is taken to be a single frequency component, and given by the expression:

$$(n\bar{\omega}_j) = e^{i\nu t} e^{i\nu S/c} \sum_{n=0}^{\infty} \frac{A^{(n)}_j}{(i\nu/c)^n} \qquad (28)$$

The function $S(\bar{x})$, sometimes called the eikonal, defines geometrical wave-fronts or wave surfaces for which $\nu$ is the frequency of the wave and $S$ is a constant. Geometrical optics and the eikonal equation can be derived from the lowest order terms of (28). In particular, to define a local wavefront we choose large frequency components $\nu \to \infty$ so that our scale of interest contains many wavelengths and we are insured that the lowest order terms of (28) capture most of the physics.

To proceed we plug the lowest order term of (28) into our wave equation in the form (25), and take the limit of infinite frequency. This gives:

$$1 = (\delta_{jk} + \frac{\eta_{jk}}{c^2}) \frac{\partial^2 S}{\partial x_j \partial x_k} \qquad (29)$$

Multiplying equation (29) by the spacelike distance squared and equating the eikonal with our choice of metric $dS = cdt$ gives the form:

$$dx_i dx_l \delta_{il} = (c^2 \delta_{jk} + \eta_{jk}) \frac{dt^2 dx_i dx_l \delta_{il}}{dx_j dx_k} \qquad (30)$$

Because our choice of physical metric forces this to be true in any inertial reference frame, the difference between the terms on either side of (30) must vanish (for light) in any frame, and has the correct units for the assignment:

$$ds^2 = dx_i dx_l \delta_{il} - (c^2 \delta_{jk} + \eta_{jk}) \frac{dt^2 dx_i dx_l \delta_{il}}{dx_j dx_k} \qquad (30)$$

which is consistent with (26).

In this manner we can build our space-time geometry from the equations governing light in subspace. Although the method as applied here was for a simple subspace, and many assumptions went into the specific wave equation used, the method should work in more general cases. For example, we have considered only equations of state independent of polarization and in media described by distribution functions isotropic in peculiar spin; changing this choice will likely require a more general approach to defining our physical geometry. However, even our simple choice of perturbations around a flat-space geometry gives enough freedom to describe many commonly studied geometries.

In the simple model as developed, we have allowed freedom only in the nine component subspace pressure tensor. The Einstein equations give ten degrees of freedom for the line element. However, almost never are geometries considered that make use of all these degrees of freedom. It is not clear at this stage whether the missing degree of freedom is a sign of problem or a prediction for this simple model.

## IV. Applications and Implications

*In which future consequences of the atomistic space-time on the frontiers of general relativity are hypothesized.*

Although the derivation of the form of the quasi-metric from moments of an atomistic transport equation is compelling and an aid to visualizing physics fundamentals, we have not made any predictions other than those which were already formulated many decades ago. However, as more detailed theories are advanced in this arena, from more complex subspaces and specific equations of state, deviations from the classical GR approach will emerge. The new physics will of course appear at the frontiers of the discipline, where the experimental (observational) evidence is thin and the approximations of classical GR no longer hold. Such systems are generally referred to as the realm of quantum gravity, the implication being that more complete theory unifying aspects of these two different disciplines is required. Here we briefly examine a few of these areas, again applying our simple model as an example.

### A. The Structure of Matter

The last century has seen incredible progress in understanding the nature of matter. The theory of elemental atoms was incredibly successful, and was improved upon by nuclear hadronic physics, and finally by the introduction of quarks in quantum chromodynamics. Of course, the motivation that has driven physicists thus far is still there, and the same questions can still be asked of the latest "fundamental particles", e.g. quarks and leptons. Phenomena such as electron – positron annihilation and flavor changing suggest that leptons are not strong atoms but do have constituent parts.

From the kinetic space-time viewpoint, a successful theory will model these particles as coherent structures in subspace. This task will not be attempted here, but we can discuss a few aspects of a solution (and the lack thereof) qualitatively. A model of a lepton will reproduce the electric field, magnetic moment, and spin, and so will involve some kind of vortex dynamics. It will also require quantization of mass (for the three species of lepton) which suggests soliton structure. These models will also need to be consistent with the Dirac equation and quantum electrodynamics, as well as generate the necessary perturbation of the subspace pressure tensor to account for the role of mass in Einstein's equation, which links the stress-energy tensor of matter to the quasi-metric of space-time.

One promising step in this task was taken by [Madelung, 1926], who found that the Schrödinger equation could be transformed into basic hydrodynamic transport equations. This was approach was also used to model a kinetic space-time model by [Winterberg, 1970 & 1995], whose approach is similar to that of this research, but for quantum systems. More recent work along these lines has showed a statistical interpretation for quantum operators and wave-functions consistent with the approach of second atomization [Kaniadakis, 2002].

Quantized particles can act as sources of light (photoelectric effect). It therefore appears that the collision terms in the wave equation (15) will be important in describing macroscopic (in this context) particles, such as quarks and leptons, as well as more general anisotropic distributions. Unfortunately nucleon models will be more complex still. It is important to note that attempts toward understanding the nature of matter near such a system as a black hole are unlikely to succeed if we cannot model ordinary matter, e.g. the mass ratio of the proton and the electron!

## B. Black Holes

For a simplified spherically symmetric stationary uncharged star, the geometry is usually described by the Schwarzschild quasi-metric (28). In our derivation, we assumed a subspace pressure tensor of the form (27). In this simple model, the pressure near a gravitating body is thus slightly lower in the non-radial directions. Another way to view this field is that while the speed of light in physical coordinates is locally the same everywhere, the speed of light in subspace coordinates is slightly less in the non-radial directions, and slightly greater as one moves away from the source mass. This can explain the curvature of light rays e.g. near the sun, and the observed gravitational redshift.

In physical space-time coordinates, there is a singularity at the Schwarzchild radius of the black hole, when the perturbation $\eta = 1$. In subspace, this corresponds to a zero pressure in the perpendicular directions. Dynamically this seems as though it would be unstable, as the collision terms must play a role in such an unlikely arrangement of velocities. The addition of these terms, or other kinetic effects, could help resolve the physical singularity, and could be forced to reproduce theoretical quantum gravity effects in this region including Hawking radiation.

One problem in our derivation is that in (25) we assumed the perturbation to the flat space metric to have very small spatial derivatives. For the perturbation $\eta = \dfrac{2MG_N}{c^2 r}$, the only spatial derivative is $\dfrac{\partial \eta}{\partial r} = \dfrac{-2MG_N}{c^2 r^2}$, which satisfies our criteria as a small quantity for large distances $r$. However, when we approach the Schwarzschild radius ($r_s = \dfrac{2MG_N}{c^2}$) this becomes larger: $\frac{\partial \eta}{\partial r} = \frac{-c^2}{2MG_N}$. The magnitude of these two contributing factors (the perturbation and its derivative), become equal at the Schwarzschild radius for a central mass $M = c^2 / 2G_N$, about 0.0003 solar masses. In other words, for any normal black hole we need not worry about contributions of the derivative term outside the Schwarzschild radius. Nevertheless, the physicist interested in applying the kinetic theory inside the event horizon will need to re-derive the line element including these derivative terms.

## C. Cosmological Expansion

The usual model of cosmological expansion is expressed by the Friedman-Lemaître-Robertson-Walker quasi-metric:

$$ds^2 = A(t)[dr^2 + d\theta^2 + \bar{r}^2 \sin^2 \theta d\phi^2] - c^2 dt^2 \tag{31}$$

The bar over the radial coordinate indicates a 'proper motion distance' and a total curvature of the universe will change that term. We will not go into the detailed evidence for the form of the function *A(t)* and the curvature parameters here, presented in e.g. [Misner, Thorne and Wheeler, 1970; Peebles, 1993]. Recent evidence from supernovae surveys and cosmic microwave background measurements are consistent with an accelerating *A(t)* and a flat space, although inflationary theories suggest different histories.

From the form of our quasi-metric (30) as derived from a perturbation of flat space, equation (31) suggests a cosmological equation of state:

$$n \langle \omega_i V_j V_k \rangle = (n \bar{\omega}_i) \delta_{jk} \frac{c^2}{A(t)} \tag{32}$$

For the case of constant density this is equivalent to a form of the subspace pressure tensor:

$$\langle V_i V_j \rangle = \delta_{ij} \frac{c^2}{A(t)} \tag{33}$$

Here an expanding universe (*A(t)* increasing with time) is interpreted as a cooling subspace distribution. A big bang (or big crunch) singularity which occurs as $A(t) \to 0$ suggests a distribution in which the subspace atoms move with near infinite speed, and the speed of light in subspace coordinates grows without bound. It is not necessary to use a non-Euclidean subspace geometry to reproduce these results.

## V. Summary

The approach has been speculative, but the methodology consistent. An atomic kinetic model in subspace is first chosen. From this model, a preferred wave-form is chosen, and a physical metric defined. Anisotropies in the distribution of atoms alter the propagation of the chosen wave-forms, and give a dynamic physical space-time. This procedure was carried out here for a simple model of Euclidean subspace, and shown to be consistent with many features of general relativity.

However, more questions are raised than answered, for it appears the field is in its infancy. How can we use the atomic theory as presented to model more macroscopic particles such as the quark and the electron, maintaining consistency with GR? What groups of subspace geometries will accurately model our physical space-time? And what utility can knowledge of subspace dynamics be for dwellers of physical space-time? It is hoped that the answers to these and other questions of second atomization will prove as exciting and fruitful to 21$^{st}$ century physics as the chemical atoms did for 20$^{th}$ century physics.